\documentclass[twocolumn,aps,floatfix,prl]{revtex4}
\usepackage{amssymb}
\usepackage{amsmath}
\usepackage{graphicx}

\begin{document}

\title{RSFQ devices with selective dissipation for quantum information processing}
\author{J. Hassel, H. Sepp\"a and P. Helist\"o, }
\affiliation{VTT Information Technology, Microsensing, P.O. Box 1207, 02044 VTT, Finland}

\begin{abstract}
We study the possibility to use frequency dependent damping in RSFQ circuits
as means to reduce dissipation and consequent decoherence in RSFQ/qubit 
circuits. We show that stable RSFQ operation can be achieved by shunting 
the Josephson junctions with an $RC$ circuit instead of a plain resistor. 
We derive criteria for the stability of such an arrangement, and discuss the 
effect on decoherence and the optimisation issues.
We also design a simple flux generator aimed at manipulating flux qubits. 
\end{abstract}

\maketitle

\bigskip

Rapid single flux quantum (RSFQ) technology \cite{lik1} has been suggested as the
classical interface for the quantum bits \cite{sem1}, and eventually for a scalable
quantum computer. RSFQ technology is inherently dissipative. The
dissipation is a likely source of decoherence, which limits the allowed
coupling between the RSFQ circuit and the quantum circuit. It is caused by
the damping of the Josephson junctions by shunt resistors. The conventional
damping is, however, higher than what is needed for stable operation.
Therefore one is encouraged to search solutions to decrease it. One approach
is to use nonlinear damping in order to switch the damping on only, as a
junction is switching \cite{zor1}. Our approach, on the other hand, is based
on the fact that the switching events occur at the time scale of the inverse
plasma frequency. Therefore the damping at lower frequencies is redundant.
The simplest way to realise the high-pass filtering is to connect a
capacitor in series with the shunt resistor. One benefit is that such a
circuit is realisable by a conventional Nb/AlOx trilayer process \cite{gro1}. 
A similar
approach has previously been suggested and tested to produce low-noise SQUID
magnetometers \cite{sep1,kiv1}, and as means to improve the resolution of 
flux qubit readout circuits \cite{rob1}. We now show that it is also possible 
to realise generic full-scale RSFQ circuits with such a configuration. 
As an example, we introduce a device design able to drive a qubit into 
a coherent superposition of flux states. The effect on the
decoherence is also discussed.

In simple terms, the stability of an RSFQ circuit is guaranteed by the
sufficient damping of the plasma resonances of the junctions and of the $LC$
resonances formed by inductors and junction capacitances. The maximum (zero
bias) angular plasma frequency is $\omega_{p}=1/\sqrt{L_{J}C}$, where $L_{J}=\Phi
_{0}/2\pi I_{c}$ is the Josephson inductance, $C$ the capacitance 
and $I_{c}$ the critical current of junction, and $\Phi _{0}=h/2e$ 
is the flux quantum. Therefore the Q-value of
the plasma resonance is chosen below unity. For a conventional damping
scheme (see Fig. 1(a) and (b)), the square of the Q-value is given by the
Stewart-McCumber parameter $\beta _{c}=2\pi I_{c}R_{s}^{2}C/\Phi _{0}$,
where $R_{s}$ is the shunt resistance and $\Phi _{0}=h/2e$ is the flux
quantum. The inductances of the RSFQ circuit elements are of the same order
as the Josephson inductance (or $\beta _{L}\equiv 2\pi LI_{c}/\Phi _{0}\sim
1 $), so this simultaneously ensures the damping of the $LC$ resonances.

\begin{figure}[b]
\includegraphics[width=7cm]{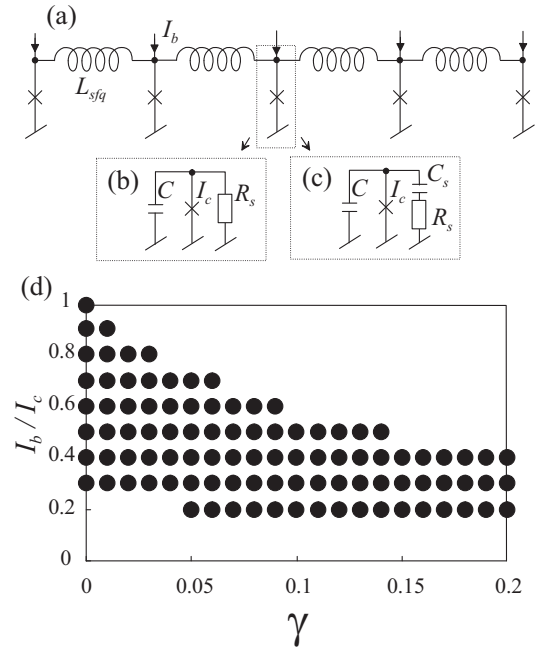}
\caption{(a) A Josephson transmission line realised with (b) conventional and (c) frequency
dependent damping. (d) Stable parameter range as function of $\gamma = 1/R_s C_s \omega_p$
and the bias current $I_b$ scaled to critical current $I_c$.}
\end{figure}

The junction parameters, inductances, and the shunt resistance can be
similarly defined and their parameters chosen
for the $RC$ shunted RSFQ (Fig. 1(c)) as well. The additional component
value
to be chosen is the shunt capacitance $C_{s}$. From the discussion
above it follows that a natural additional stability parameter 
is the ratio of the $%
R_{s}C_{s}$ cutoff and the plasma frequency, namely 
$\gamma =1/\omega_{p}R_{s}C_{s}$. 
To test the effect of $\gamma $ on the stability, we
simulate the most basic RSFQ element, the Josephson transmission line (JTL).
The value of $\beta _{c}=1/2$ is fixed, while the bias point $I_{b}$ and
the capacitance $C_{s}$ (or $\gamma $) are varied in order to test the
stable range of parameters. We define the system to be stable, if the flux
quantum propagates from the left end to the right end correctly as shown in
Fig. 2(b). The indications of the lost stability are error pulses (in practice,
the flux quantum reflecting back from the right end) or junctions switching
permanently into a finite voltage state. The resulting stable parameter
range is shown in Fig. 1(d). The leftmost column shows the
corresponding result with the conventional JTL (formally with $C_{s}=\infty $%
). The decreased stability with $RC$ damping
and large bias currents is mainly because the potential barrier
protecting against the error pulses is reduced. However,
with realistic values of $C_{s}$ sufficient stability can be obtained.

In a practical realisation of the shunt capacitance it is important to avoid
parasitic resonances. The wavelength at the plasma frequency in the
capacitor is given as $\lambda_p = 2\pi c/\omega_{p}\sqrt{\varepsilon _{r}\left(
1+2\lambda _{L}/d\right)} $, where $c$ is the speed of light, $\varepsilon
_{r}$ is the dielectric constant, $\lambda _{L}$ is the London penetration
depth of the electrodes, and $d$ is the insulator thickness. To be
on the safe side, the dimension of the capacitor should be $\lambda_p /8$ at
maximum. Therefore for the capacitance (of a square) it applies $%
C_{s}\lesssim (\pi c)^{2}\varepsilon _{0}/16\omega_{p}^{2}d\left( 1+2\lambda
_{L}/d\right)$. In other words, realizability dictates that

\begin{equation}
\gamma  \gtrsim \frac{16 \omega_{p}d\left( 1+2\lambda _{L}/d\right) }{\pi^2
R_{s}c^{2}\varepsilon _{0}}=\frac{32\sqrt{2}d\left( 1+2\lambda _{L}/d\right)
}{\pi \Phi _{0}\varepsilon _{0}c^{2}}I_{c},  \label{amin}
\end{equation}
where in the last form the definition of the plasma frequency and $\beta
_{c}=1/2$ have been used. The minimum realizable $\gamma $
depends only on the critical current, capacitor thickness and the London
penetration depth. It is also favorable to use a small critical current,
which is in accordance to minimising the heating effects. For example, an
existing Nb process for milli-Kelvin applications has Nb$_{2}$O$_{5}$
capacitors with $d =$ 140 nm, $\lambda_L=$ 90 nm, and typically $I_{c}=3$ $\mu $A, whence it
follows $\gamma \gtrsim 0.008$ thus enabling the operation well 
in the stable regime (see Fig. 1(d)).

We test next $RC$ damping by simulating a simple 
device (Fig. 2(a)) able to generate rectangular fast rise-time 
flux-pulses. The device consists of two DC/SFQ converters
\cite{pol1} driving an RS flip-flop \cite{lik1}. The generator 
takes two periodic (e.g. sinusoidal) mutually phase-locked
signals as inputs, and produces a flux through the output 
coil $L_{10}$. The frequency of the flux pulses is 
the frequency of the input signals, and the pulse length is
related to the phase difference of them. Resulting simulated
time domain plots are shown in Figs. 2(b) and 2(c). 
The pulses with amplitude $\sim \Phi_0$ 
can e.g. be used in manipulating a flux-type 
qubit \cite{mak1}, provided the coupling between 
the device and the qubit is strong enough. With such 
a device one avoids the need of wide-band wiring and consequent 
noise from the room temperature electronics. A further benefit 
is relative simplicity. 

\begin{figure}
\includegraphics[width=8.5cm]{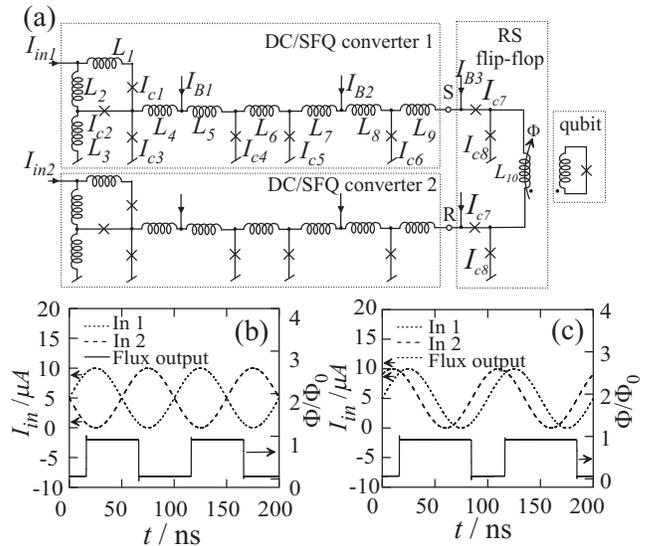}
\caption{(a) A flux pulse generator circuit coupled to a flux qubit. All the Josephson 
junctions of the RSFQ circuit are damped 
with $RC$ shunts, which are not shown for clarity. b) and (c) simulated 
time-domain plots of the RSFQ circuit. The parameters 
used here are $I_{c1}$...$I_{c7}=I_{0}$, $I_{c8}=1.4I_0$, $I_{b1}=1.6I_0$,
$I_{b2}=1.5I_0$, $I_{b3} = 0.7 I_0$, where $I_0 =$ 2.9 $\mu$A and $L_1=0.35L_0$, 
$L_2=0.33L_0$, $L_3 = L_6 = 0.6L_0$, 
$L_4=0.1L_0$, $L_5=0.3L_0$, $L_7=L_8 = 0.5L_0$, 
$L_9 = L_0$, and $L_{10}=2.5L_0$, where $L_0=$ 357 pH. In addition $\omega_p =2\pi\times$19 GHz,
$\beta_c = 0.5$ and $\gamma = 0.1$.}
\end{figure}

The effect of drive and readout circuits on quantum circuits depends largely
on the qubit type and the realisation of the classical circuit.
Here we consider in general terms qubits, whose flux degree of freedom 
\cite{cal1,moo1,chi1,vio1,plo1} is inductively coupled to an
RSFQ circuit (Fig. 3(a)). This type of an experiment benefits probably most
from the frequency dependent damping.

\begin{figure}
\includegraphics[width=7cm]{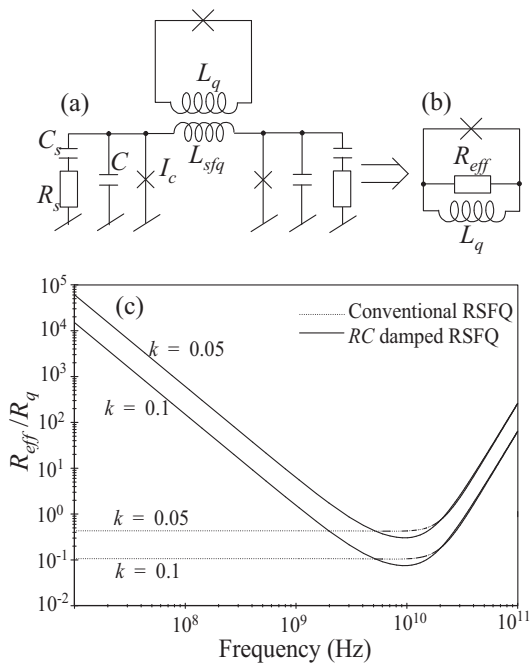}
\caption{Calculation of the effective damping resistance of a flux qubit coupled 
to an RSFQ circuit. The example parameters used here are $L_{sfq} = 357$ pH,
$L_{q} = 10$ pH, $I_c$ = 3 $\mu$A, $\beta_c=0.5$, and $\gamma=0.2$.}
\end{figure}

The dissipation can be modelled as a frequency
dependent effective resistance in parallel to the qubit 
inductance coupled to the RSFQ
circuit (Fig. 3 (b)). The effective resistance $R_{eff}$ is calculated for
both conventional and $RC$ shunted RSFQ in Fig. 3(c). For the conventional
RSFQ $R_{eff}$ is constant at low frequencies leading to constant
dissipation 
\begin{equation}
R_{eff,0}=b\frac{R_{s}}{k^{2}}\frac{L_{q}}{L_{sfq}},  \label{ref0}
\end{equation}
where $L_q$ is an inductance of the qubit, $L_{sfq}$ is the inductance 
of the RSFQ circuit, $k$ is the coupling between the two, and
$b$ depends on the details of the RSFQ circuit. Taking only the
nearest elements of the RSFQ circuit into account, we get $%
b=(1/2)(1+4(L_{J}L_{sfq}+L_{sfq}^{2})/L_{J}^{2})$, where the terms of order 
$k^{2}$ have been dropped. For the conventional RSFQ technology, the
dissipation is ohmic, i.e. the environment spectral density $J_{1}\left(
\omega \right) =\left( \pi /2\right) \alpha \hbar \omega $ \cite{cal1,leg1}. 
The decoherence time is typically inversely proportional to $J_1 (\omega_q)$,
where $\omega _{q}=\Delta E/\hbar $ with $\Delta E$ the energy 
level splitting of the qubit \cite{leg1}. For a flux type qubit the 
dissipation parameter
$\alpha =B\times R_{q}/R_{eff},_{0}$, where $R_q=h/4e^2$ is the
quantum resistance and $B\sim 1$
is a constant dependent on the qubit details \cite{mak1}. 
The minimum requirement for
coherent operation (the weak-damping limit) is that $\alpha \ll 1$. 
E.g. with parameters used in Fig. 3 this leads to the requirement of 
the coupling factor $k\ll 0.03$. This in turn leads to severe limitations 
in the resolution of a readout application, or a limited flux 
amplitude in the generation of drive signals.

In case of RC shunted RSFQ the corresponding figure is
\begin{equation}
R_{eff}\left( \omega \right) =\frac{R_{eff,0}}{(\omega R_{s}C_{s})^{2}} =
R_{eff,0}\left(\frac{\gamma \omega_p}{\omega}\right)^2.
\label{ref}
\end{equation}
This leads to superohmic spectral density $J_{2}\left( \omega \right)
=(\pi / 2)\left(\gamma \omega_p\right) ^{-2}\alpha \hbar \omega ^{3}$
\cite{leg1,rob1}, and  the improvement 
in the decoherence time (if limited by the RSFQ 
circuit)\ is $\left(\gamma \omega_p /\omega_q\right) ^{2}$, provided 
$\omega _{q}\ll \gamma \omega_p$. This enables significant increase
in $k$, even close to unity. 

To optimise an RSFQ/qubit system, $R_{eff}$ should be maximised.
Since $R_{eff} \propto \gamma^2$ (Eq. (\ref{ref})), $\gamma$ 
should be chosen as large as possible. The drawback is, though, 
that the stability against the parameter spread is decreased 
(see Fig. 1(d)). Another possibility is to 
increase the plasma frequency, i.e. increase the critical current 
density $J_c$. 
It can be shown that $R_{eff}$ is proportional to $J_c^{3/2}$,
if $L_q$, the stability parameters and $k$ are held
constant. Therefore it is favorable to use large
$J_c$, which is also favorable in terms on maximising 
the RSFQ speed. To simultaneously minimise the self-heating effects, 
one should have small $I_c$ junctions \cite{sav1}. 
Therefore large $J_c$ junctions with small areas are optimal.
However, if the area is limited by the fabrication,
one needs to compromise between the speed, the dissipation 
experienced by the qubit, and the heating effects. 

\vspace{0.5cm}
The authors wish to thank M. Kiviranta, A. Kidiyarova-Shevchenko, J. Pekola, and 
A.O. Niskanen for useful discussions. The work was supported by EU through 
project RSFQubit (no. FP6-502807).

\end{document}